\documentclass[aps,prc,twocolumn,superscriptaddress,showpacs]{revtex4}
\usepackage[dvips]{graphicx}
\newcommand{\nc}{\newcommand}           

\nc{\nuc}[2]    {$^{#1}${#2}}           
\nc{\vc}[1]     {\mbox{\boldmath $#1$}} 
\nc{\tht}        {\theta}                
\nc{\lam}       {\lambda}               
\nc{\bras}[1]   {\langle #1|}           
\nc{\kets}[1]   {|#1\rangle}            
\nc{\hO}        {\hat{O}}               
\nc{\wtil}      {\widetilde}            
\nc{\lw}[1]     {\smash{\lower1.25ex\hbox{#1}}}  
\nc{\mapleft}[1]{			
 \smash{\mathop{\,			%
  \hbox to 1.2cm{\rightarrowfill}\, }\limits_{#1}}}

\def\JL#1#2#3#4{ {{\rm #1}}\ #2(#3)#4}  
\nc{\PR}[3]     {\JL{Phys. Rev.}{#1}{#2}{#3}}
\nc{\PRC}[3]    {\JL{Phys. Rev.~C}{#1}{#2}{#3}}
\nc{\PRA}[3]    {\JL{Phys. Rev.~A}{#1}{#2}{#3}}
\nc{\PRL}[3]    {\JL{Phys. Rev. Lett.}{#1}{#2}{#3}}
\nc{\NP}[3]     {\JL{Nucl. Phys.}{#1}{#2}{#3}}
\nc{\PL}[3]     {\JL{Phys. Lett.}{#1}{#2}{#3}}
\nc{\PTP}[3]    {\JL{Prog. Theor. Phys.}{#1}{#2}{#3}}
\nc{\PTPS}[3]   {\JL{Prog. Theor. Phys. Suppl.}{#1}{#2}{#3}}
\nc{\PRep}[3]   {\JL{Phys. Rep.}{#1}{#2}{#3}}
\nc{\ZP}[3]     {\JL{Z. Phys.}{#1}{#2}{#3}}
\nc{\JP}[3]     {\JL{J. of Phys.}{#1}{#2}{#3}}
\nc{\andvol}[3] {{\it ibid.}\JL{}{#1}{#2}{#3}}

\begin{document}
\title{
Three-body Coulomb breakup of $^{11}$Li in the complex scaling method
}

\author{Takayuki Myo\footnote{Corresponding author.\\ 
Postal address: Research Center for Nuclear Physics (RCNP), Osaka University, Ibaraki 567-0047, Japan\\
Tel/Fax: +81-6-6879-8940 / +81-6-6879-8898\\
E-mail address: myo@rcnp.osaka-u.ac.jp (T.Myo)}
}
\affiliation{
Research Center for Nuclear Physics (RCNP),
Ibaraki, Osaka 567-0047, Japan}%

\author{Shigeyoshi Aoyama}
\affiliation{
Information Processing Center, Kitami Institute of Technology,
Kitami 090-8507, Japan.
}

\author{Kiyoshi Kat\=o}
\affiliation{
Division of Physics, Graduate School of Science,
Hokkaido University, Sapporo 060-0810, Japan.
}

\author{Kiyomi Ikeda}
\affiliation{
RI-Beam Science Laboratory, RIKEN (The Institute of Physical and Chemical Research), 
Wako, Saitama 351-0198, Japan.
}

\date{\today}

\begin{abstract}
Coulomb breakup strengths of \nuc{11}{Li} into a three-body \nuc{9}{Li}+$n$+$n$ system
are studied in the complex scaling method.
We decompose the transition strengths into the contributions from three-body resonances, 
two-body ``\nuc{10}{Li}+$n$'' and three-body ``\nuc{9}{Li}+$n$+$n$'' continuum states. 
In the calculated results, we cannot find the dipole resonances with a sharp decay width in \nuc{11}{Li}. 
There is a low energy enhancement in the breakup strength, 
which is produced by both the two- and three-body continuum states.
The enhancement given by the three-body continuum states is found to have
a strong connection to the halo structure of \nuc{11}{Li}.
The calculated breakup strength distribution is compared with the experimental data from MSU, RIKEN and GSI.
\end{abstract}


\maketitle 


Studies of unstable nuclei have obtained much attention
with the development of radioactive beams\cite{Ta96}.
The \nuc{11}{Li} nucleus is known as a typical Borromean system,
in which the \nuc{9}{Li}+$n$ and $n$+$n$ subsystems
do not have any bound states, but the total \nuc{9}{Li}+$n$+$n$ system has
a bound state.
This Borromean mechanism is considered to play an important role in the formation of a halo,
but it is not yet fully understood.
The observed large matter radius of \nuc{11}{Li}, 
which is an evidence of the halo, suggests
a large mixing of the $(1s_{1/2})^2$ neutron component in addition to the $(0p_{1/2})^2$ one. 
Another interesting problem related to the halo structure of \nuc{11}{Li}
is a characteristic property of the excitation mode.
For the excited states of \nuc{11}{Li},
the so-called soft dipole resonance\cite{Ha87,Ike92} is 
expected in the low-energy region.
In the shell model picture, the major component of the soft dipole
resonance is described as $(1s_{1/2})(0p_{1/2})$.
Thus, the behavior of the $1s$- and $0p$-orbits of valence neutrons is very crucial
to understand the halo structure and the excited states in \nuc{11}{Li}.


Experimentally, measurements of the Coulomb breakup strength distributions of \nuc{11}{Li} 
have been carried out by three groups at MSU\cite{Ie93}, RIKEN\cite{Sh95} and GSI\cite{Zi97}. 
The low energy enhancement of the strength seems to indicate the existence
of the soft dipole resonance, although the shapes of distributions 
obtained by the three experiments at different incident energies do not 
coincide with each other.
In addition to these measurements, observations of the two-body correlation
provide us with a key to discuss the mechanism of the breakup reaction.
The measured invariant mass spectrum of \nuc{9}{Li}+$n$ 
shows the low energy enhancement\cite{Zi97,Ko93}.
This result implies the existence of the low-lying $1s$-orbit
near the \nuc{9}{Li}+$n$ threshold energy in \nuc{10}{Li}. Such a $1s$-orbit in the \nuc{9}{Li}-$n$ system 
is expected to provide ``a continuum structure'' in the three-body breakup reaction of \nuc{11}{Li}. 
On the other hand, in the case of the two-body breakup reaction of \nuc{11}{Be},
the low energy enhancement observed in the strength is understood
as ``a continuum response" of a large low-momentum component in the halo structure
of the ground state\cite{Na94}. 
In order to understand the observed enhancement of the \nuc{11}{Li} breakup strength, 
therefore, we must consider the continuum structure of
the \nuc{10}{Li}+$n$ binary component and the continuum response
of \nuc{9}{Li}+$n$+$n$ in addition to the three-body resonance.


Theoretically, many methods such as the Faddeev method, the
hyperspherical harmonics approach and sophisticated variational methods 
have been developed to solve the Borromean systems\cite{Zu93}.
However, there is a discrepancy between some theoretical results for the soft dipole resonances in \nuc{11}{Li}. 
Garrido {\em et al.}\cite{Ga02b} calculated the dipole strength distribution,
and predicted at least three dipole resonances, 
but we did not obtain any resonant solution with a sharp width ($\Gamma/2<$E$_r$) in the variational method\cite{Mu98,Ao02}. 
This discrepancy is considered to come from a difference on the \nuc{9}{Li}-$n$ potential.
The dipole strength distribution calculated by Garrido {\em et al.} does not agree with the observed strengths 
within the experimental ambiguity.
The peak energy is slightly lower and the width is much sharper than the experimental one.
Thus, it is strongly desirable to obtain detailed information about the
\nuc{9}{Li}-$n$ potential, the excited resonant states and the continuum
responses through analyses of the Coulomb breakup reaction of \nuc{11}{Li}.


For this purpose, we have been developing the applicability of the complex scaling method (CSM)\cite{ABC} 
which has recently received much attention for finding three-body resonances\cite{Ao02,Ga02b}.
It is a big advantage of CSM that for an unbound system conduces to 
the separation not only between resonances and continuum states 
but also between different kinds of continuum states starting from different thresholds\cite{My01}.
This advantage of CSM is exploited in the calculation of transition strengths of unbound states 
beyond the two-body systems. 
We showed a successful result for the three-body Coulomb breakup reaction 
of a simpler two-neutron halo nucleus \nuc{6}{He}\cite{My01}. 
The results for \nuc{6}{He} are  summarized as (i) no resonance peak 
corresponding to the soft dipole mode is obtained, and (ii) the \nuc{5}{He}($3/2^-$)+$n$ binary component 
dominates the $E1$ strength distribution and the responses of the other components, such as three-body continuum states 
of \nuc{4}{He}+$n$+$n$, are very small. 

In this letter, we extend this method to \nuc{11}{Li} and investigate the continuum structures and responses
through the Coulomb breakup reaction. 
We briefly explain an extended \nuc{9}{Li}+$n$+$n$ model for \nuc{11}{Li}, 
and report the results of the Coulomb breakup strength distributions comparing with experimental data. 
From the obtained results, we discuss the mechanism of the breakup processes. 


We describe \nuc{11}{Li} with an extended \nuc{9}{Li}+$n$+$n$ three-body model\cite{My02}. 
The Hamiltonian of this model is given in the orthogonality condition model\cite{Sa77} as follows:
\begin{eqnarray}
	H(\mbox{\nuc{11}{Li}})
&=&	H(\mbox{\nuc{9}{Li}})
+	\sum_{i=1}^3{t_i} - T_G
+	\sum_{i=1}^2V_{cn,i} + V_{nn} 
\nonumber
\\
&&\hspace{1cm}+~\lambda_{PF}\, \kets{\phi_{PF}}\bras{\phi_{PF}},\
	\label{eq:Ham}
\end{eqnarray}
where $H(\mbox{\nuc{9}{Li}})$, $t_i$ and $T_G$ are 
the internal Hamiltonian of \nuc{9}{Li},
the kinetic energy of each cluster and the center-of-mass of the three-body system, respectively.
The \nuc{9}{Li}-$n$ potential, $V_{c n}$, is given by a folding-type one with MHN interaction\cite{Fu80}.
For the potential $V_{nn}$ for two valence neutrons, the Minnesota potential\cite{Ta78} is used with parameter 
$u$=0.95.
The last term in Eq.~(\ref{eq:Ham}) is a projection operator 
to remove the Pauli forbidden (PF) states from the \nuc{9}{Li}-$n$ relative motion\cite{Ku86}.
The definition of the PF states is also given in Ref.~\cite{My02},
where the value of $\lambda_{PF}$ is taken as $10^6$~MeV in the present calculation.

The folding-type \nuc{9}{Li}-$n$ potential was originally constructed 
so as to produce energy splittings in the \nuc{10}{Li} spectra, 
such as $1^+$--$2^+$ (for the $p_{1/2}$-neutron) and $1^-$--$2^-$ (for the $s_{1/2}$-neutron) 
due to the coupling between spins of the valence neutron and \nuc{9}{Li}($3/2$)\cite{Ka93}.
However, in the study of \nuc{11}{Li}\cite{My02}, we discussed to add 
a phenomenological tail potential to the original folding-type potential
to improve the behaviour of the tail part of the \nuc{9}{Li}-$n$ potential.
The behavior of the $s$-wave state near the threshold is very 
sensitive to the tail part of the potential due to the spatial extension of the wave function.
The tail potential also plays an important role in lowering the energy of the $(1s_{1/2})^2$ component 
with respect to that of the $(0p_{1/2})^2$-component in \nuc{11}{Li}(\nuc{9}{Li}+$n$+$n$).
Then, we have two parameters in the \nuc{9}{Li}-$n$ potential; 
one is the $\delta$ to change the strength of the second range of the
folding part, and another the strength of phenomenological tail potential given by a Yukawa-form\cite{My02}.
They are determined so as to reproduce the $1^+$ resonance of
\nuc{10}{Li} at 0.42 MeV\cite{Ka93} and the $s$-wave property which is a virtual state showing a large negative scattering length.


The wave function of \nuc{11}{Li} is given as 
\begin{eqnarray}
	\Psi^J(^{11}{\rm Li})
&=&	\sum_{i}^{N_c}{\cal A}\left\{\, [\Phi^{3/2^-}(C_i), \chi^{j}_i(nn)]^J\,
	\right\}.
	\label{eq:WF_11Li}
\end{eqnarray}
Here, the \nuc{9}{Li} nucleus is expressed by a multi-configuration
$\sum_i a_i\, \Phi^{3/2^-}(C_i)$ in order to take into account the neutron pairing correlation\cite{Ka99}. 
Since the mixing amplitude $a_{i}$ depend on the relative distances
between \nuc{9}{Li} and two valence neutrons in the \nuc{9}{Li}+$n$+$n$
system, we express it by the function $\chi^{j}_{i}(nn)$. 
We should notice that $\chi^{j}_{i}(nn)\rightarrow a_i\times(\mbox{plane wave})$ at large distances 
between \nuc{9}{Li} and two valence neutrons. 
For the total spin $J$ of \nuc{11}{Li}, $j$ expresses the spin of two valence neutrons, and then $J=3/2\otimes j$.

We describe the wave function of valence neutrons using the combined set
of the two kinds of the basis states; the cluster orbital shell model (COSM; V-type)
and the extended cluster model (ECM; T-type),
which we call as the hybrid-TV model\cite{Ike92,Mu98,To90b}:
\begin{eqnarray}
    \chi^{j}_i(nn)
&=&   \chi^{j}_{i,V}(\vc{\xi}_V)+\chi^{j}_{i,T}(\vc{\xi}_T),
\end{eqnarray}
where $\vc{\xi}_V$ and $\vc{\xi}_T$ are V-type and T-type coordinate sets, respectively.
The radial part of the relative wave function is expanded 
with a finite number of Gaussian basis functions centered at the origin.

By employing $C_1=(0p_{3/2})^4_\nu$ and $C_2=(0p_{3/2})^2_\nu(0p_{1/2})^2_\nu$
for $p$-shell neutrons in \nuc{9}{Li} ($N_c$ is 2),
we solve a coupled-channel \nuc{9}{Li}+$n$+$n$ three-body problem. 
We use the MHN interaction\cite{Fu80} to calculate the neutron pairing correlation in \nuc{9}{Li}, 
which leads to the $|a_2|^2=15~\%$ of the pairing excitation\cite{My02}. 
When the valence neutrons approach to \nuc{9}{Li}, 
the coupling between the motion of the valence neutrons and 
the pairing correlation in \nuc{9}{Li} becomes stronger. 
It was discussed that this dynamical coupling in
\nuc{10}{Li}(\nuc{9}{Li}+$n$) provides the so-called pairing-blocking 
effect\cite{Ka99} which explains the lowering of $1s$-orbits in \nuc{10}{Li}. 
For \nuc{11}{Li}, we adjust the $(0p)^2$-$(1s)^2$ pairing coupling
between valence neutrons to reproduce the observed binding energy of \nuc{11}{Li} (0.31MeV)\cite{Au93}.

\begin{table}[t]
\caption{
 Results for the three types of wave functions of the present model;
 (upper) Scattering lengths $a_s$ and energies $E$ of the virtual state
 ($2^-$,$1^-$) of \nuc{10}{Li}, (lower) $(1s_{1/2})^2$ probability $P$ and matter radius $R_m$ of the
 \nuc{11}{Li} ground state.
}\label{tab:model}
\begin{center}
\begin{ruledtabular}
\begin{tabular}{c|cccc}
\nuc{10}{Li}               & P-1   & P-2  & P-3  \\
\noalign{\hrule height 0.5pt}
$a_s(2^-)$ [fm]            & $-$12.7 & $-$17.0  & $-$21.7 \\
$E(2^-)$ [MeV]             & $-$0.05 & $-$0.03  & $-$0.02 \\
\noalign{\hrule height 0.25pt}
$a_s(1^-)$ [fm]            & $-$6.5 & $-$8.6  & $-$10.7   \\
$E(1^-)$ [MeV]             & $-$0.08 & $-$0.06  & $-$0.05 \\
\end{tabular}

\begin{tabular}{c|cccc}
\nuc{11}{Li}               & P-1   & P-2  & P-3  & exp. \\
\noalign{\hrule height 0.5pt}
P[$(1s_{1/2})^2$] [\%]      &  21.0  & 29.4 & 38.8 & --- \\ 
\lw{$R_m$ [fm]}            & \lw{3.33}  & \lw{3.58} & \lw{3.85} & 3.12$\pm$0.16\protect$^a$ \\
                           &     &      &      & 3.53$\pm$0.06\protect$^b$ \\
\end{tabular}
\end{ruledtabular}
\end{center}
\vspace{-0.3cm}
\noindent
$^a$Reference\cite{Ta88b},~$^b$Reference\cite{To97}
\end{table}

In the analysis of excited states in \nuc{11}{Li}, we prepare the three types of
the \nuc{11}{Li} wave function; P-1, P-2 and P-3, 
which are characterized by the $(1s_{1/2})^2$ probability in the ground state.
In Table~\ref{tab:model}, we list the properties of the prepared
\nuc{10}{Li} and \nuc{11}{Li} wave functions.
The $s$-orbit properties are different among them,
which affect on the scattering length of the \nuc{10}{Li} $s$-state and 
the size of the halo structure of the \nuc{11}{Li} ground state.
We also pay attention to the effects of the halo structure on the breakup strength. 
They can be seen from the responses of resonances and continuum states in \nuc{11}{Li}.

\begin{figure}[th]
\centering
\includegraphics[width=6.5cm,clip]{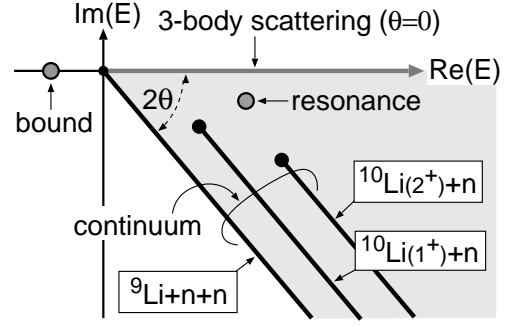}
\caption[]{
Schematic distribution of the energy eigenvalues of the \nuc{9}{Li}+$n$+$n$ system 
with CSM. 
}
\label{fig:ABC}
\end{figure}

We calculate the unbound states of \nuc{11}{Li} applying CSM to the extended \nuc{9}{Li}+$n$+$n$ system,
where the relative coordinates between \nuc{9}{Li} and the two neutrons are transformed with a scaling angle $\theta$ as
\begin{eqnarray}
	\vc{\xi}_{V,T}
&\to&	\vc{\xi}_{V,T}\, e^{i\theta}\, .
\end{eqnarray}
The momenta corresponding to the asymptotic channel $\alpha$ of \nuc{9}{Li}+$n$+$n$ and \nuc{10}{Li}$^{(*)}$+$n$
are also transformed as
\begin{eqnarray}
	{\bf k}_\alpha
&\to&	{\bf k}_\alpha\, e^{-i\theta}\, .
\end{eqnarray}
Here, we notice that $2\theta$ corresponds to a rotation angle of the cuts in the Riemann sheets of complex energies,
and the angular part of the wave function does not change in CSM.
In Fig.~\ref{fig:ABC}, we show a schematic energy eigenvalue distribution 
of the complex-scaled \nuc{9}{Li}+$n$+$n$ system governed by the ABC-theorem\cite{ABC}.
When $\theta=0$, (unbound) scattering states are obtained on the real energy axis, which
includes all components of resonances and continuum states.
For a finite value of $\theta$, 
the continuum states are obtained on the Riemann cuts rotated down by $2\theta$, 
and hereafter we call these rotated continuum states as the continuum ones. 
When we take a large $\theta$, as shown in Fig.~\ref{fig:ABC}, in addition to the three-body bound
states (3BB), we obtain (i) discrete three-body resonances (3BR), 
(ii) two-body continuum states (2BC) of \nuc{10}{Li}(1$^+$, 2$^+$)+$n$, 
and (iii) three-body continuum states (3BC) of \nuc{9}{Li}+$n$+$n$,
which are decomposed from the three-body scattering states. 
Two-body continuum states of \nuc{10}{Li}+$n$ are expressed by the two straight lines
whose origins are resonance positions of \nuc{10}{Li}($1^+,~2^+$).

Since the virtual states of \nuc{10}{Li}($1^-,~2^-$) and of 2$n$ cannot be
located in CSM due to the limitation of the scaling angle $\theta$, 
the channels of \nuc{10}{Li}($1^-,~2^-$)+$n$ and \nuc{9}{Li}-2$n$ components
are included in 3BC of \nuc{9}{Li}+$n$+$n$.

Using the Green's function, the strength function for the operator ~$\hO_\lam$ with rank $\lam$ is expressed as 
\begin{eqnarray}
	{\cal S}_\lam(E) 
&=&	-\frac1{\pi}\ {\rm Im}
	\biggl[
	\int d\vc{\xi} d\vc{\xi}'               	\:
	\wtil{\Psi}_i^*(\vc{\xi})\: \hO^\dagger_\lam 	\:
	{\cal G}(E,\vc{\xi},\vc{\xi}')              	\:
	\nonumber\\
&\times&\hO_\lam \Psi_i(\vc{\xi}')
	\biggr]\, ,
	\label{eq:strength0}
\end{eqnarray}
where $\Psi_i(\vc{\xi})$ is the initial wave function of \nuc{11}{Li}.
We apply the complex scaling to the right hand side of Eq.~(\ref{eq:strength0}):
\begin{eqnarray}
	{\cal S}_\lam(E)
&=&	-\frac1{\pi}\ {\rm Im}
	\biggl[
	\int d\vc{\xi} d\vc{\xi}'              		\:
	[\wtil{\Psi}_i^*(\vc{\xi})]^\tht (\hO^\dagger_\lam)^\tht\:
	{\cal G}^\tht(E,\vc{\xi},\vc{\xi}')              \:
\nonumber\\
& & \hspace{2cm}\times\hO_\lam^\tht \Psi^\tht_i(\vc{\xi}')	\biggr],
	\label{eq:strength1}
\end{eqnarray}
where the complex-scaled Green's function ${\cal G}^\tht(E,\vc{\xi},\vc{\xi}')$ is given as
\begin{eqnarray}
{\cal G}^\tht(E,\vc{\xi},\vc{\xi}')
&=&	\left\langle \vc{\xi} \left| \frac{ {\bf 1} }{ E-H(\tht) } \right| \vc{\xi}' \right\rangle\, ,
	\label{eq:green0}
	\\
&=&	\sum_\nu\hspace*{-0.5cm}\int\
	\frac{\Psi^\tht_\nu(\vc{\xi}) [\wtil{\Psi}^*_\nu(\vc{\xi})]^\tht}{E-E_\nu^\tht}
= 	\sum_\nu\hspace*{-0.5cm}\int\
	{\cal G}^\tht_\nu(E,\vc{\xi},\vc{\xi}')\, .
	\label{eq:green1}
\end{eqnarray}
In this expansion, $E_\nu^\theta$ and $\Psi^\tht_\nu(\vc{\xi})$ ($\wtil{\Psi}^\tht_\nu(\vc{\xi})$) are 
the energy eigenvalues and eigenfunctions (bi-orthogonal eigenfunctions\cite{My01,Be68,My98}) of 
the complex-scaled Hamiltonian $H(\theta)$, respectively. 
Therefore, summation and/or integration are taken over $\nu$ of the solutions of $H(\tht)$ 
including 3BR, 2BC of \nuc{10}{Li}($1^+$,$2^+$)+$n$ and 3BC of \nuc{9}{Li}+$n$+$n$ 
(There is no bound state except for the ground state).

Inserting the complex-scaled Green's function in Eq.~(\ref{eq:strength1}),
we obtain the strength function decomposed into each component ${\cal S}_{\lam,\nu}(E)$ for the final state $\nu$ as
\begin{eqnarray}
{\cal S}_\lam(E)
&=&	\sum_\nu\hspace*{-0.5cm}\int\ {\cal S}_{\lam,\nu}(E)\, .
	\label{eq:strength2}
	\\
	{\cal S}_{\lam,\nu}(E)
&\equiv&-\frac1{\pi}\ {\rm Im}
	\left[	 \frac{
		\bras{\wtil{\Psi}_i^\tht}  (\hO^\dagger_\lam)^\tht \kets{\Psi_\nu^\tht}
		\bras{\wtil{\Psi}_\nu^\tht} \hO_\lam^\tht          \kets{\Psi_i^\tht}
	  }{E-E_\nu^\tht}\right]\, .
	\label{eq:strength}
\end{eqnarray}
It is noted that the total strength ${\cal S}_{\lam}(E)$ is an observable and positive definite for any energy
and independent of $\theta$.
On the other hand, the partial strengths ${\cal S}_{\lam,\nu}(E)$ of resonance and continuum components
are not necessarily positive definite, and in fact show sometimes negative values.
A detailed explanation is discussed in Refs.\cite{My01,My98}.
Since the number of resonances obtained in the calculation depends on $\theta$ and 
jumps at a certain $\theta$, ${\cal S}_{\lam,\nu}(E)$ has a discontinuity at some $\theta$ values\cite{My98}. 

In the present calculation, we solve the eigenvalue problem of the complex-scaled Hamiltonian with the hybrid-TV model,
and the discretized approximation is adopted for continuum states. 
We use 25 Gaussian basis functions for one relative motion and set their maximum range about 40 fm.
We numerically checked the reliability of the discretized representation
of continuum states and the stability of the calculated strengths by changing the parameters of the basis functions. 

\begin{figure}[b]
\centering
\includegraphics[width=8.5cm,clip]{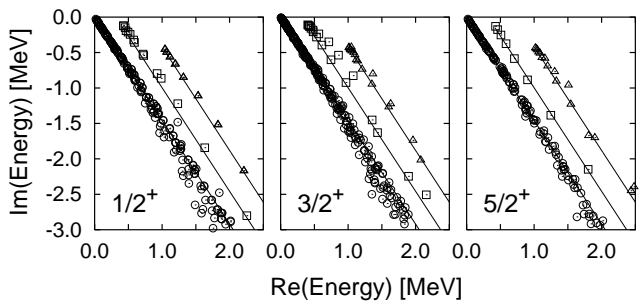}
\caption{
Energy eigenvalues of three dipole excited states ($1/2^+$, $3/2^+$, $5/2^+$) of \nuc{11}{Li}
with $\theta$=28 degree in CSM.
Squares and triangles indicate 2BC of \nuc{10}{Li}($1^+$)+$n$ and \nuc{10}{Li}($2^+$)+$n$, respectively.
Circles indicate 3BC of \nuc{9}{Li}+$n$+$n$.} 
\label{fig:ene_1-}
\end{figure}

In Fig.~\ref{fig:ene_1-}, we show the eigenvalue distributions of $1/2^+$, $3/2^+$ and $5/2^+$ states
in a complex energy plane at $\theta=28$ degrees.
These are dipole excited states ($j=1$) of \nuc{11}{Li} from the ground state ($J^\pi=3/2^-$). 
We obtain all eigenvalues along three lines of rotated Riemann cuts corresponding to 
two 2BC of \nuc{10}{Li}($1^+$,$2^+$)+$n$ and one 3BC of \nuc{9}{Li}+$n$+$n$ as discussed in Fig.~\ref{fig:ABC}.
There is no dipole resonance which is located between 
the real energy axis and the rotated continuum lines of the excited states.
In other words, we cannot find any resonances with a sharp width at least ($\Gamma/2E_r<\tan^{-1}2\theta$;~$\theta=28^\circ$).
This result means that the dipole strengths are exhausted by continuum states of \nuc{11}{Li}.
Since the wave function of $s$-wave valence neutrons is spatially extended,
dipole resonances including $s$-wave components in \nuc{11}{Li} may tend to decay easily, 
and so resonances have large decay widths, whose poles are located below the rotated continuum states. 
The effect of such resonances is included in the continuum spectra in this calculation.

\begin{figure}[b]
\centering
\includegraphics[width=6.5cm,clip]{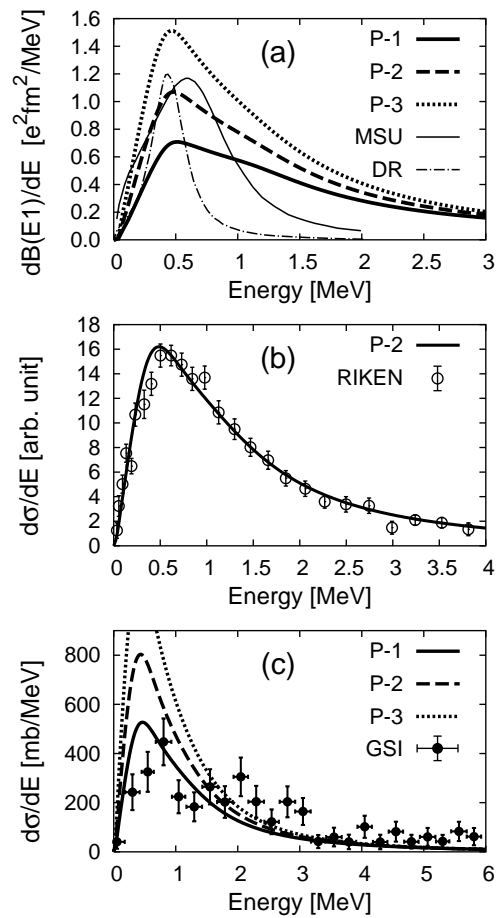}
\caption{
Calculated dipole strengths and cross sections of \nuc{11}{Li} 
in comparison to the theory (denoted as DR, Fig.~5 (a) of Ref.~\cite{Ga02b}) in (a) 
and the experimental data;(a)\cite{Ie93}, (b)\cite{Sh95} and (c)\cite{Zi97},
where we take into account the convolution.}
\label{fig:cross}
\end{figure}

Using the solutions of the continuum spectra of the dipole excited states, 
we calculate the dipole transition from the ground state.
In Fig.~\ref{fig:cross}(a), we show the results of the dipole strength
functions summing up the contributions from the $1/2^+$, $3/2^+$ and $5/2^+$ states 
for the three types of the ground state wave function.
The energy is measured from the \nuc{9}{Li}+$n$+$n$ threshold.
It is found that the strengths show the low energy enhancement 
whose height is sensitive to the $(1s_{1/2})^2$ probability of the \nuc{11}{Li} ground state. 
This enhancement is interpreted as a threshold effect coming from 
the continuum states and reflects the halo structure of \nuc{11}{Li}.
We compare our results to the experimental data of MSU\cite{Ie93}.
The position of the enhancement almost agrees with the data, but
a disagreement of the shape is seen in the strength above the energy 1 MeV.
We also compare our results with the calculation (denoted as DR in Fig.~\ref{fig:cross}(a)) by Garrido {\em et al.}\cite{Ga02b}. 

In Figs.~\ref{fig:cross}(b) and \ref{fig:cross}(c),
we derive the cross sections of the \nuc{11}{Li} breakup 
by multiplying the transition strength and the virtual photon number in the equivalent photon method\cite{Be88},
where the target is Pb in both cases.
We can see a good agreement with the data of RIKEN\cite{Sh95} for the P-2 wave function with the $(1s_{1/2})^2$
probability being around $30\%$. However, the magnitude of the cross section is not observed.
It would be desirable to determine its magnitude.
For the data of GSI\cite{Zi97}, the P-1 wave function with the $(1s_{1/2})^2$ probability being around $20\%$ gives a good result. 
However, the experimental error bars are still large. 
Further experimental data with high resolution and statistics would be required.

\begin{figure}[b]
   \centering
   \includegraphics[width=6.5cm,clip]{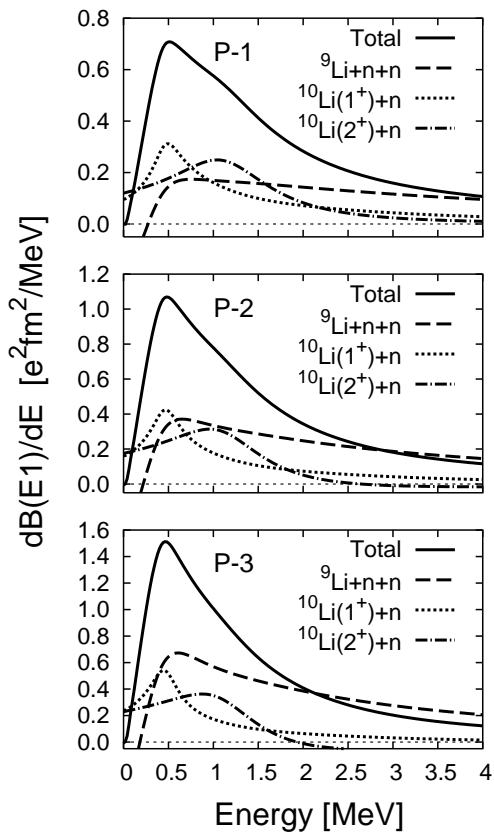}
\caption{Decomposition of the dipole transition strength of \nuc{11}{Li}
with the three types of the $(1s_{1/2})^2$ probability in the ground state,
as described in the text.}
\label{fig:strength2}
\end{figure}


In Fig.~\ref{fig:strength2}, we show the result of the separation of the $E1$ transition strength of \nuc{11}{Li}
into two-body and three-body continuum components for the three types of the $(1s_{1/2})^2$ probability in the ground state.
It is found that the two-body continuum component of \nuc{10}{Li}($1^+$)+$n$ shows a low energy enhancement in each panel, 
whose peak position is just above the two-body threshold (0.42 MeV) of \nuc{10}{Li}($1^+$)+$n$.
Another two-body continuum component of \nuc{10}{Li}($2^+$)+$n$ shows 
a broader structure because a larger decay width of the $2^+$ state 
than that of the $1^+$ state in \nuc{10}{Li} broadens the strength. 
The transition strengths into the \nuc{10}{Li}(1$^+$,2$^+$)+$n$ 2BC
correspond to the following two kinds of physical situations:
First one is that one of the valence neutrons of $(0p_{1/2})^2$ in the \nuc{11}{Li} ground state is excited
to a continuum state of $s$- or $d$-waves by the $E1$ external field, 
and a remaining $p$-orbital valence neutron forms resonances of \nuc{10}{Li} with \nuc{9}{Li}.
This situation is similar to the case of \nuc{6}{He} breakup reactions\cite{My01}
because the \nuc{6}{He} ground state is almost dominated by $(0p_{3/2})^2$ of two valence neutrons.
Second one is that one of the valence neutrons of $(1s_{1/2})^2$ in the \nuc{11}{Li} ground state is excited
to a $p$-orbit and forms resonances of \nuc{10}{Li}(1$^+$,2$^+$) with \nuc{9}{Li}, 
and a remaining $s$-orbital valence neutron becomes a continuum state.
This is the characteristic situation of \nuc{11}{Li}.
The contribution of this process can be seen as the differences 
between the strengths of 2BC of P-1, P-2, P-3 in Fig.~\ref{fig:strength2}.
The low energy enhancement in two-body continuum components is interpreted as a threshold effect 
of 2BC of the \nuc{10}{Li}($1^+$,$2^+$)+($s$-wave neutron) channels.

\begin{table}[b]
\caption{Integrated strengths in comparison to NEW-SRV and the ratios
of the each component of the strength for three types of the $(1s_{1/2})^2$ probability.}
\label{tab:NEW-SRV}
\begin{center}
\begin{ruledtabular}
\begin{tabular}{l|c|cccc}
\multicolumn{2}{l|}{}                         &  P-1  &  P-2 &  P-3     \\
\noalign{\hrule height 0.5pt}
\multicolumn{2}{c|}{Integral [e$^2$fm$^2$] }  &  1.401 &  1.901  &  2.439 \\
\multicolumn{2}{c|}{NEW-SRV [e$^2$fm$^2$]  }  &  2.215 &  2.856  &  3.593 \\
\noalign{\hrule height 0.25pt}
      & \nuc{9}{Li}+$n$+$n$    ~~~~~&  0.392 &  0.522  &  0.658 \\
~~Ratios~~~& \nuc{10}{Li}($1^+$)+$n$~~~~~~&  0.306 &  0.262  &  0.216 \\
      & \nuc{10}{Li}($2^+$)+$n$~~~~~~&  0.302 &  0.216  &  0.126 \\
\end{tabular}
\end{ruledtabular}
\end{center}
\end{table}


From Fig.~\ref{fig:strength2}, it is also found that the contribution from 3BC of \nuc{9}{Li}+$n$+$n$ 
increases as the $(1s_{1/2})^2$ probability increases.
The strength distribution of 3BC has a peak at the low energy region ($\sim$ 0.5 MeV), 
and slowly decreases with energies. At energies higher than about 2 MeV, the strength of 3BC becomes dominant.
The difference between the 3BC and 2BC strength distributions may be understood from 
the level density of the continuum states.
The level density of 3BC is widely distributed, because 3BC have two
degrees of freedom of the relative motions, but 2BC has one degree of freedom. 

With the increasing of the $(1s_{1/2})^2$ probability in the \nuc{11}{Li} ground state, 
the magnitude of the strengths of 2BC increases, but the shapes do not change much.
On the other hand, the strength of 3BC shows a sharper peak at a low energy, 
and its magnitude increases markedly with the $(1s_{1/2})^2$ probability in comparison to the case of 2BC.
This result indicates that the structure of 3BC depends strongly on the halo structure of the ground state.
To see it more clearly,
we evaluate the integrated strengths up to 5 MeV for the three types of $(1s_{1/2})^2$ probabilities,
which exhaust about $65\%$ of the non-energy weighted sum rule value
(NEW-SRV) as shown in Table~\ref{tab:NEW-SRV}.
We also list the ratios of the each component of 2BC and 3BC to the total integrated strength.
It is found that the ratios of the contribution of 3BC increase proportionally with the $(1s_{1/2})^2$ probability. 

These results mean that the components of 2BC and 3BC, contribute comparably in the Coulomb breakup reaction of \nuc{11}{Li},
and that, in particular, 3BC depends strongly on the halo structure of the ground state.
In fact, a large contribution of 3BC cannot be seen in the case of \nuc{6}{He} breakup reactions\cite{My01}.
In the three-body model of \nuc{6}{He}, the $(1s_{1/2})^2$ probability of two valence neutrons is 2.4\% 
which is much smaller than that of \nuc{11}{Li}.
These results indicate that the mechanisms of the breakup reactions of \nuc{6}{He} and \nuc{11}{Li} are different,
and this fact is caused by the different $(1s_{1/2})^2$ probabilities in their ground states.

There may be broad resonances including virtual states which are located
under the Riemann cuts of 2BC and 3BC calculated here. 
However, their effects are considered to produce no remarkable structure, 
although they are seen to give a considerable strength.
In the present calculation, we cannot conclude how much they could contribute to the strength.


In summary, we investigate the three-body Coulomb breakup reaction of \nuc{11}{Li}
employing the complex scaling method (CSM)
to describe the three-body unbound states of the \nuc{9}{Li}+$n$+$n$ system.
We decompose the transition strengths into every component of the
unbound states, such as three-body resonances, and two- and three-body continuum states,
and examine the effects of each component on the strength distribution of \nuc{11}{Li}. 

From the results, we cannot find any dipole resonances with a sharp width.
The observed low energy enhancement in the dipole strength comes from the continuum states of \nuc{11}{Li}, 
where the \nuc{10}{Li}+$n$ two-body continuum states 
and the \nuc{9}{Li}+$n$+$n$ three-body continuum ones
give comparable contributions.
This result also means that the breakup mechanism of \nuc{11}{Li} is different from that of \nuc{6}{He}
where the two-body continuum states corresponding to the sequential breakup process are dominant in the strength.
Furthermore, it is found that the $(1s_{1/2})^2$-component in the \nuc{11}{Li} ground state
is responsible for the increases of the contribution of the three-body continuum states and 
the low energy enhancement in the strength.
This calculated Coulomb breakup strength distribution of \nuc{11}{Li} shows 
a good agreement with the experimental data of RIKEN.

As the origin of the low energy enhancement in the strength,
the threshold effect due to the continuum responses is considered,
which is seen in both components of the two- and three-body continuum states.
Here, the strength of three-body continuum states include a contribution from the virtual states in \nuc{10}{Li},
which we do not consider in this study.
It is interesting to investigate the effect of virtual states on the breakup reaction in the future works.

\begin{acknowledgments}
The authors would like to acknowledge valuable discussions with 
Professor S. Shimoura. 
This work is supported by a Grant-in-Aid for Scientific Research (No.~12640246)
of the Ministry Education, Culture, Sports, Science and Technology, Japan.
One of the authors (T.~M.) thanks to the Japan Society for the Promotion of Science (JSPS) for support.
Computational calculations of this work are supported by Hokkaido University Computing Center (HUCC).
This work was performed as a part of the ``Research Project for Study of
Unstable Nuclei from Nuclear Cluster Aspects (SUNNCA)'' sponsored by RIKEN.
\end{acknowledgments}


\end{document}